\documentclass{ws-procs9x6}
\begin{document}

\title{
Hybrid Quantum Annealing for Clustering Problems
}

\author{Shu Tanaka}
\address{
Research Center for Quantum Computing, 
Interdisciplinary Graduate School of Science and Engineering, Kinki University, 
3-4-1 Kowakae, Higashi-Osaka, Osaka 577-8502, Japan\\
E-mail: shu-t@chem.s.u-tokyo.ac.jp
}

\author{Ryo Tamura}
\address{
Institute for Solid State Physics, University of Tokyo,
5-1-5 Kashiwanoha, Kashiwa, Chiba 277-0805, Japan\\
E-mail: r.tamura@issp.u-tokyo.ac.jp
}

\author{Issei Sato}
\address{
Information Science and Technology, University of Tokyo 
7-3-1, Hongo, Bunkyo-ku, Tokyo 113-0033, Japan\\
E-mail: sato@r.dl.itc.u-tokyo.ac.jp
}

\author{Kenichi Kurihara}

\address{
Google,
6-10-1, Roppongi, Minato-ku, Tokyo 106-0032, Japan\\
}

\begin{abstract}
We develop a hybrid type of quantum annealing in which we control temperature and quantum field simultaneously.
We study the efficiency of proposed quantum annealing and find a good schedule of changing thermal fluctuation and quantum fluctuation.
In this paper, we focus on clustering problems which are important topics in information science and engineering.
We obtain the better solution of the clustering problem than the standard simulated annealing by proposed quantum annealing.
\end{abstract}


\bodymatter

\section{Introduction}\label{sec:intro}

Optimization problems have spread in wide area of science, for example, information science and statistical physics.
Properties of optimization problems can be summarized as follows:
\begin{itemize}
 \item There are many elements.
 \item A cost function can be defined.
       The best solution is the state where the cost function takes the maximum value or the minimum value depending on the definition of the given problem.
\end{itemize}
An well-known example of optimization problem is the so-called ``traveling salesman problem''.
The traveling salesman problem is to find the shortest path covering all the cities.
Here, we can path through each city only once.
In this problem, the cost function is the length of path.
In general, it is difficult to obtain the best solution by naive approach since the number of candidates of a solution is very huge.
Then, in the optimization problem, it is a central issue how to find the best (or better) solution.
There are many types of optimization problems.
Depending on an optimization problem, there are specific algorithms to solve the problem.

To treat optimization problem in a general way, on the other hand, a couple of approaches have been developed from a viewpoint of statistical physics.
The cost function can be regarded as the internal energy in statistical physics.
The cost function of the best solution corresponds to the internal energy of the ground state.
Kirkpatrick {\it et al.} proposed a pioneering generic algorithm -- simulated annealing\cite{Kirkpatrick-1983,Kirkpatrick-1984}.
Simulated annealing is a method to obtain not so bad solution of optimization problem by decreasing temperature.
In high temperature, since the probability distribution at equilibrium state is almost flat, the state can be changed easily.
As we decrease temperature gradually, generated probability distribution expects to approach the equilibrium probability distribution at each temperature.
Then we can find not so bad solution of problem by simulated annealing.
Because of the Geman-Geman's argument\cite{Geman-1984}, we can succeed to obtain the best solution with the probability unity if we decrease the temperature slow enough.
Although there could be better algorithms specific to each optimization problem, simulated annealing is regarded as a stable and generic method because of easy implementation and independency from problems.
The purpose of our study is to establish a good general method for optimization problems from a viewpoint of quantum statistical physics.

The organization of this paper is as follows.
We first review on quantum annealing which has been believed a good general method to obtain not so bad solution of optimization problems.
In section 3, we will review how to implement Monte Carlo simulation.
In section 4, we will introduce a model which can treat clustering problems.
We also consider quantum effect on this model in this section.
In section 5, some remarks on this problem will be shown.
In section 6, we will show results obtained by our proposed method.
In section 7, we will summarize our study.
We will review on the concepts of ``invisible fluctuation'' in the appendix A.

\section{Quantum annealing}\label{sec:qa}

Quantum annealing is expected that it succeeds to obtain the better solution of optimization problems than simulated annealing\cite{Kadowaki-1998,Kadowaki-2002,Farhi-2001,Santoro-2002,Charkrabarti-book1,Charkrabarti-book2,Das-2008,Ohzeki-2011}.
This method is based on quantum statistical physics.
There have been a couple of realization methods of quantum annealing: (i) stochastic method, (ii) deterministic method, and (iii) experiment on artificial lattice.

The stochastic method is realized by quantum Monte Carlo method in many circumstances\cite{Santoro-2002,Kurihara-2009}.
The quantum Monte Carlo method is an established method to obtain equilibrium properties of strongly correlated quantum systems.
Efficient algorithms for quantum annealing have been developed such as the cluster algorithm\cite{Nakamura-2008,Morita-2009}.
Owing to these masterly methods, quantum Monte Carlo simulation can be adopted for large-scale systems.

There are a couple of deterministic methods for quantum annealing.
The first one is based on the time-dependent Schr\"odinger equation\cite{Kadowaki-1998}.
This method can trace the real-time evolution which can be observed in real experiments.
This method is called as ``quantum adiabatic evolution'' in quantum information science\cite{Farhi-2001}.
This method cannot treat large-scale systems because of the limitation of size of memory in computer.
The second one is based on a time-dependent density matrix renormalization group\cite{Suzuki-2007,Laguna-2007}.
By this method, we can study time-evolution of one-dimensional quantum systems.
However it is difficult to treat two-dimensional or three-dimensional quantum systems by this method.
Both of two methods are based on the principle of quantum mechanics.
However, it is not necessary to treat the Schr\"odinger equation directly if we adopt the deterministic method as quantum annealing.
This is because our purpose is to obtain not so bad solution of given problem.
One of the examples is mean-field calculation\cite{Tanaka-2002,Sato-2009}.
This method can treat large-scale systems as well as the quantum Monte Carlo simulation.
Then, this method has been widely adopted for optimization problems.

Next there are a couple of proposals for experiments on artificial lattice for quantum annealing.
For example, we can generate quantum state by optical lattice\cite{Jane-2003,Porras-2004,Lewenstein-2007,Friedenauer-2008}.
In addition, the Ising model with transverse field can be realized by superconducting flux qubits\cite{Harris-2009,Berkley-2010,Harris-2010,Amin-2011}.
These methods are expected as a new type of quantum computer.

There have been a number of studies on quantum annealing from a viewpoint of theoretical physics.
Convergence theorem for quantum annealing was proved as well as that for simulated annealing\cite{Morita-2006,Morita-2007,Morita-2008}.
According to this theorem, the permitted upper bound of sweeping speed of quantum field in quantum annealing is larger than that of temperature in simulated annealing.
From this theorem, quantum annealing seems better than simulated annealing in principle.
However, we often sweep temperature and/or quantum field faster than that upper bound in practice.
Then, it is nontrivial whether quantum annealing is better than simulated annealing from a viewpoint of practical situation.
Microscopic behavior in quantum annealing has been also investigated.
It is well-known that frustrated systems have some interesting static behavior induced by thermal fluctuation and quantum fluctuation.
Dynamical properties of frustrated systems have been studied from a viewpoint of quantum annealing\cite{shu-t-2007a,Matsuda-2009,shu-t-2009a,shu-t-2010a,shu-t-2010b,shu-t-2010d}.
Novel type of implementation of quantum annealing itself is an important topic.
One of the examples is quantum annealing based on the Jarzynski equality\cite{Ohzeki-2010a}.
This method is expected an efficient method since it uses both merits which come from thermal fluctuation and quantum fluctuation.

In this paper, we adopt the first strategy -- the quantum Monte Carlo method as the realization of quantum annealing.

\section{Monte Carlo simulation}\label{sec:mc}

In this section, we review how to implement Monte Carlo simulation.
It has been often used in order to obtain the equilibrium properties of strongly correlated systems such as magnetic systems and bosonic systems.
Equilibrium physical quantities of the system which is expressed by the Hamiltonian ${\cal H}$ at finite temperature $T$ is given as 
\begin{eqnarray}
 \langle {\cal O} \rangle_{\rm eq}^{(T)} = 
  \frac{{\rm Tr}\, {\cal O} {\rm e}^{-\beta {\cal H}}}{{\rm Tr}\, {\rm e}^{-\beta {\cal H}}},
\end{eqnarray}
where $\beta$ denotes the inverse temperature $1/T$ and here the Boltzmann constant $k_{\rm B}$ is set to be unity.
If we consider a small system, we can obtain all of the equilibrium physical quantities by naive method.
If we consider large scale systems, however, we cannot obtain the equilibrium properties by naive method in practice.
Monte Carlo simulation enables us to calculate equilibrium physical quantities with high accuracy by the following relation.
\begin{eqnarray}
 \frac{\sum_\Sigma {\cal O}(\Sigma) {\rm e}^{-\beta {\cal H}(\Sigma)}}{\sum_\Sigma {\rm e}^{-\beta {\cal H}(\Sigma)}}
  \to \langle {\cal O} \rangle_{\rm eq}^{(T)},
\end{eqnarray}
where $\Sigma$ denotes sample, in other words, state.
Physical quantity converges to the equilibrium value as the number of samples increases.
In fact, the above calculation is inefficient if states are generated by uniform distribution.
In order to make the method more efficient, we generate states according to equilibrium probability distribution which is proportional to the Boltzmann factor ${\rm e}^{-\beta E(\Sigma)}$.
This method is called the importance sampling.
We can obtain the equilibrium value as follows:
\begin{eqnarray}
 \frac{\sum_\Sigma {\cal O}(\Sigma)}{\sum_\Sigma} \to \langle {\cal O} \rangle_{\rm eq}^{(T)}.
\end{eqnarray}
In order to generate a state from the equilibrium distribution, we just have to use Markov chain Monte Carlo method.
Time evolution of probability distribution is given as the master equation:
\begin{eqnarray}
 \nonumber
 P(\Sigma_i,t+\Delta t) = 
  &&-\sum_{j \neq i} P(\Sigma_i,t) w(\Sigma_j|\Sigma_i) \Delta t \\
  &&+ \sum_{j \neq i} P(\Sigma_j, t) w (\Sigma_i|\Sigma_j) \Delta t
  + P(\Sigma_i,t) w(\Sigma_i|\Sigma_i) \Delta t,
\end{eqnarray}
where $P(\Sigma_i,t)$ denotes probability of the state $\Sigma_i$ at time $t$ and $w(\Sigma_j|\Sigma_i)$ represents transition probability from the state $\Sigma_i$ to the state $\Sigma_j$ in unit time.
Transition probability $w(\Sigma_j|\Sigma_i)$ obeys
\begin{eqnarray}
 \sum_{\Sigma_j} w(\Sigma_j|\Sigma_i) = 1 \,\,\,\, (\forall \Sigma_i).
\end{eqnarray}
The master equation can be represented as 
\begin{eqnarray}
 {\bf P}(t+\Delta t) = {\cal L} {\bf P}(t),
\end{eqnarray}
where ${\bf P}(t)$ is a vector-representation of probability distribution $\{P(\Sigma_i,t)\}$ and ${\cal L}$ is called the transition matrix whose elements are expressed as
\begin{eqnarray}
 {\cal L}_{ji} = w(\Sigma_j|\Sigma_i) \Delta t,
  \,\,\,\,\,
 {\cal L}_{ii} = 1- \sum_{j\neq i}{\cal L}_{ji} = 
 1 - \sum_{j\neq i} w(\Sigma_j|\Sigma_i) \Delta t.
\end{eqnarray}
It should be noted that ${\cal L}$ is a non-negative matrix by the definition.
This time evolution is the Markovian since the time-evolution operator ${\cal L}$ does not depend on time.
If the time-evolution operator ${\cal L}$ obeys (i) detailed balance condition and (ii) ergordicity, we can obtain the equilibrium probability distribution in the limit of $t \to \infty$ because of the Perron-Frobenius theorem.

\section{Model}\label{sec:model}

Clustering problem is one of the important problems in information science and engineering.
Since it is difficult to obtain the best solution of the clustering problem by naive method, development of a new method which can obtain the best (or not so bad) solution is an important issue.
We can obtain not so bad solution by using simulated annealing as we mentioned in the section 1.
We propose a new type of quantum annealing and succeed to obtain better solution by proposed quantum annealing method as will be mentioned.

In the beginning of this section, we will explain a model to consider clustering problems.
After that, we will introduce new kind of fluctuation -- quantum fluctuation -- into this model.
Next, we will review on implementation method of the quantum annealing method.

\subsection{Clustering problem}

\begin{figure}[ht]
 \begin{center}
  $\begin{array}{ccc}
  \includegraphics[scale=0.8]{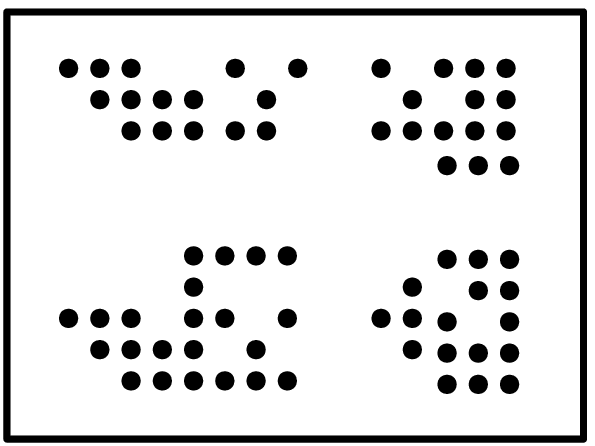}&
   \hspace{5mm}&
  \includegraphics[scale=0.8]{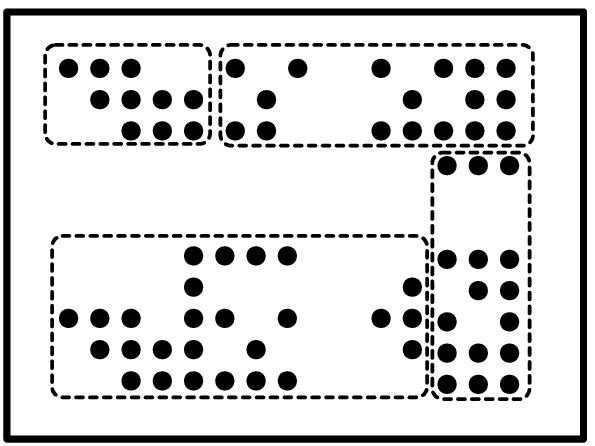} \\
    ({\rm a}) & \hspace{5mm} & ({\rm b}) \\
  \includegraphics[scale=0.8]{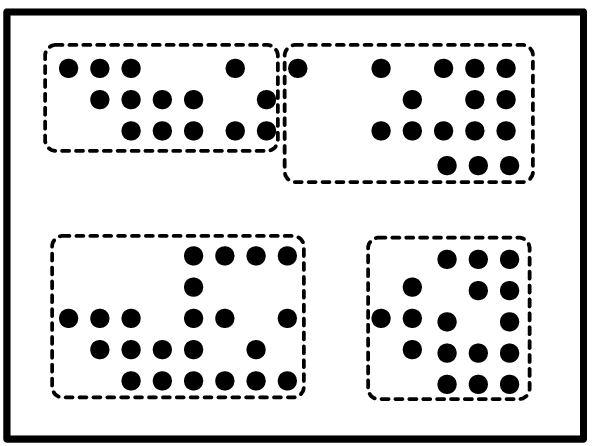}&
   \hspace{5mm}&
  \includegraphics[scale=0.8]{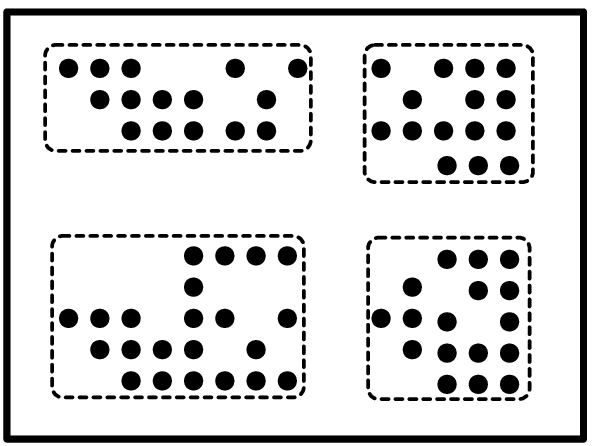} \\
    ({\rm c}) & \hspace{5mm} & ({\rm d})
  \end{array}$
  \caption{
  The dots represent elements.
  The dotted boxes denote clusters.
  In these figures, the number of elements $N=69$ and the number of clusters $Q=4$.
  (a) There are $N$ elements in the space. 
  (b) Not so good solution.
  (c) Not so bad solution.
  (d) The best solution.
  }
  \label{fig:clustering_problem}
 \end{center}
\end{figure}

In clustering problems, there are $N$ elements in the space, which is depicted in Fig.~\ref{fig:clustering_problem}(a).

Clustering problems are to decide which the best partition of these elements.
In other words, clustering problems are to search the best division of $N$ elements into $Q$ sub-categories.
Clustering problems have been applied for not only natural science but also social science.
For instance, divisions of the articles in newspaper in terms of contents and analysis of questionnaire can be considered as clustering problems.
Figure \ref{fig:clustering_problem} (b), (c), and (d) represent not so good, not so bad, and the best solution, respectively.
In practice, it is difficult to obtain the best solution by direct method for huge number $N$, since the number of total states is $Q^N$.
Then we often use simulated annealing to solve clustering problems as one of useful generic methods.
However, even if we use simulated annealing, it is difficult to obtain the best solution.
The energy landscape of clustering problems is very complicated such as random spin systems and frustrated systems.
Then we should develop more efficient algorithm than the simulated annealing method.
Actually, there is a pioneering method to obtain more better solution than simulated annealing method.
This method is called exchange method\cite{Hukushima-1996}.
In the exchange method, we prepare some independent layers where the temperatures are different.
We sometimes exchange the states between two layers according to the Boltzmann weight.
In this paper, we adopt another strategy, quantum annealing method as more efficient algorithm.
We introduce quantum term into this model to implement the quantum annealing in the next section.

\subsection{Quantum fluctuation}

Before introducing a quantum fluctuation, we review on a classical Hamiltonian of the original clustering problem.
The classical Hamiltonian is given by 
\begin{eqnarray}
 {\cal H}_{\rm c} = {\rm diag}\left( E(\Sigma_1),E(\Sigma_2),\cdots,E(\Sigma_{Q^N})\right),
\end{eqnarray}
where $E(\Sigma_i)$ denotes the eigenenergy of $i$-th state $\Sigma_i$.
Suppose we consider the case for $Q=3$ and $N=2$ as an example. 
The classical Hamiltonian is given as
\begin{eqnarray}
 {\cal H}_{\rm c} = 
  \left(
   \begin{array}{ccccccccc}
    E(\Sigma_1) & 0 & 0 & 0 & 0 & 0 & 0 & 0 & 0\\
    0 & E(\Sigma_2) & 0 & 0 & 0 & 0 & 0 & 0 & 0\\
    0 & 0 & E(\Sigma_3) & 0 & 0 & 0 & 0 & 0 & 0\\
    0 & 0 & 0 & E(\Sigma_4) & 0 & 0 & 0 & 0 & 0\\
    0 & 0 & 0 & 0 & E(\Sigma_5) & 0 & 0 & 0 & 0\\
    0 & 0 & 0 & 0 & 0 & E(\Sigma_6) & 0 & 0 & 0\\
    0 & 0 & 0 & 0 & 0 & 0 & E(\Sigma_7) & 0 & 0\\
    0 & 0 & 0 & 0 & 0 & 0 & 0 & E(\Sigma_8) & 0\\
    0 & 0 & 0 & 0 & 0 & 0 & 0 & 0 & E(\Sigma_9)
   \end{array}
  \right),
\end{eqnarray}
where the definitions of the states from $\Sigma_1$ to $\Sigma_9$ are summarized in Table \ref{table:def_sigma_q3n2}.
$\sigma_1$ and $\sigma_2$ in Table \ref{table:def_sigma_q3n2} represent the states of the first element and the second element, respectively.

\begin{table}[h]
 \tbl{Definitions of the states from $\Sigma_1$ to $\Sigma_9$.}
  {\begin{tabular}{cccccccccc}
    \toprule
    & $\Sigma_1$ & $\Sigma_2$ & $\Sigma_3$ & $\Sigma_4$ & $\Sigma_5$ & $\Sigma_6$ & $\Sigma_7$ & $\Sigma_8$ & $\Sigma_9$ \\
    \hline
    $\sigma_1$ & $1$ & $2$ & $3$ & $1$ & $2$ & $3$ & $1$ & $2$ & $3$ \\
    $\sigma_2$ & $1$ & $1$ & $1$ & $2$ & $2$ & $2$ & $3$ & $3$ & $3$ \\
    \botrule
  \end{tabular}}
 \label{table:def_sigma_q3n2}
\end{table}

Next we introduce a quantum fluctuation into this model.
Since when $Q=2$, the classical Hamiltonian is equivalent to that of the Ising model.
It is natural to extend transverse field in the Ising spin system as a quantum fluctuation.
Then we adopt the following definition as a quantum fluctuation.
\begin{eqnarray}
 {\cal H}_{\rm q} = - \Gamma \sum_{i=1}^N \sigma_i^x 
  = -\Gamma \sum_{i=1}^N (\hat{1}_Q - \mathbb{E}_Q),
\end{eqnarray}
where $\hat{1}_Q$ denotes the matrix whose all elements are unity and $\mathbb{E}_Q$ represents identity matrix.
Both $\hat{1}_Q$ and $\mathbb{E}_Q$ are $Q\times Q$ matrices.
It should be noted that $\sigma_i^x$ denotes the $x$-component of $s=1/2$ Pauli matrix at the site $i$ when $Q=2$.
Suppose we consider the case for $Q=3$ and $N=2$ as the previous example, the quantum part of the Hamiltonian ${\cal H}_{\rm q}$ is given as
\begin{eqnarray}
 {\cal H}_{\rm q} = 
 \left(
  \begin{array}{ccccccccc}
   0 & -\Gamma & -\Gamma & -\Gamma & 0 & 0 & -\Gamma & 0 & 0 \\
   -\Gamma & 0 & -\Gamma & 0 & -\Gamma & 0 & 0 & -\Gamma & 0 \\
   -\Gamma & -\Gamma & 0 & 0 & 0 & -\Gamma & 0 & 0 & -\Gamma \\
   -\Gamma & 0 & 0 & 0 & -\Gamma & -\Gamma & -\Gamma & 0 & 0 \\
   0 & -\Gamma & 0 & -\Gamma & 0 & -\Gamma & 0 & -\Gamma & 0 \\
   0 & 0 & -\Gamma & -\Gamma & -\Gamma & 0 & 0 & 0 & -\Gamma \\
   -\Gamma & 0 & 0 & -\Gamma & 0 & 0 & 0 & -\Gamma & -\Gamma \\
   0 & -\Gamma & 0 & 0 & -\Gamma & 0 & -\Gamma & 0 & -\Gamma \\
   0 & 0 & -\Gamma & 0 & 0 & -\Gamma & -\Gamma & -\Gamma & 0 
  \end{array}
  \right).
\end{eqnarray}
When we use quantum Monte Carlo method, all we have to do is to calculate the probability of the state $\Sigma$.
In the next section, we will show how to calculate the probability of the state $\Sigma$ by path-integral representation\cite{Trotter-1959,Suzuki-1976}.

\subsection{Path integral representation}

We consider the following Hamiltonian
\begin{eqnarray}
 {\cal H} = {\cal H}_{\rm c} + {\cal H}_{\rm q}.
\end{eqnarray}
When the Hamiltonian ${\cal H}$ is a diagonal matrix which corresponds to simulated annealing, {\it i.e.} $\Gamma = 0$, the probability of the state $\Sigma$ at finite temperature $T$ is given as
\begin{eqnarray}
 \label{eq:p_sa}
 p_{\rm SA}(\Sigma;\beta) = 
  \frac{{\rm e}^{-\beta E(\Sigma)}}{{\rm Tr}\, {\rm e}^{-\beta {\cal H_{\rm c}}}}
  = \frac{1}{Z}\langle \Sigma | {\rm e}^{- \beta {\cal H}_{\rm c}} | \Sigma \rangle,
\end{eqnarray}
where the denominator is called the partition function in statistical physics.
The partition function $Z$ is calculated as
\begin{eqnarray}
 Z = {\rm Tr}\, {\rm e}^{-\beta {\cal H}_{\rm c}} 
  = \sum_\Sigma \langle \Sigma | {\rm e}^{-\beta {\cal H}_{\rm c}} | \Sigma \rangle
		 = \sum_\Sigma {\rm e}^{-\beta E(\Sigma)}.
\end{eqnarray}
We can change the state by using the ``single-spin-flip'' type of heat bath method,
\begin{eqnarray}
 \label{eq:sa-update}
 p_{\rm SA}^{\rm update} (\sigma_i = s | \Sigma \backslash \sigma_i)
  = \frac{{\rm e}^{-\beta E(\sigma_i=s,\Sigma \backslash \sigma_i)}}{\sum_{s'=1}^Q {\rm e}^{-\beta E(s',\Sigma \backslash \sigma_i)}},
\end{eqnarray}
where $\Sigma \backslash \sigma_i$ means $\{\sigma_j|j\neq i\}$ and $p(A|B)$ denotes a conditional probability of $A$ given $B$.
The denominator of Eq.~(\ref{eq:sa-update}) can be calculated where the computational cost is ${\cal O}(Q)$.

In similar with the classical case, the probability of the state $\Sigma$ at finite temperature $T$ and finite quantum field $\Gamma$ is given as
\begin{eqnarray}
 \label{eq:prob_qa_direct}
 p_{\rm QA}(\Sigma;\beta,\Gamma) = 
  \frac{\langle \Sigma | {\rm e}^{-\beta {\cal H}} | \Sigma \rangle}
  {\sum_{\Sigma'} \langle \Sigma' | {\rm e}^{-\beta {\cal H}} | \Sigma' \rangle}
  = \frac{\langle \Sigma | {\rm e}^{-\beta {\cal H}} | \Sigma \rangle}{Z}.
\end{eqnarray}
Note that it is difficult to calculate $\langle \Sigma | {\rm e}^{-\beta {\cal H}} | \Sigma \rangle$, since the Hamiltonian including quantum field has off-diagonal elements.
For small systems, we can exactly calculate all of the elements ${\rm e}^{-\beta {\cal H}}$ by using the unitary transform.
However, we can not obtain them for large system in practice.
In order to calculate the probability given by Eq.~(\ref{eq:prob_qa_direct}), we should rewrite the numerator of it by path-integral representation.
Then we obtain the probability
\begin{eqnarray}
 \nonumber
 &&p_{\rm QA}(\Sigma;\beta,\Gamma) = 
  \frac{1}{Z}\langle \Sigma | 
  \left( {\rm e}^{-\frac{\beta}{m}{\cal H}_{\rm c}} {\rm e}^{-\frac{\beta}{m}{\cal H}_{\rm q}} \right)^m | \Sigma \rangle + {\cal O}(\frac{1}{m})\\
  &&= \frac{1}{Z} \sum_{\Sigma^{(1)'}} \sum_{\Sigma^{(2)}} \cdots \sum_{\Sigma^{(m)'}}
   \prod_{j=1}^m
   \langle \Sigma^{(j)} | {\rm e}^{-\frac{\beta}{m} {\cal H}_{\rm c}} | \Sigma^{(j)'} \rangle
   \langle \Sigma^{(j)'} | {\rm e}^{-\frac{\beta}{m} {\cal H}_{\rm q}} | \Sigma^{(j+1)} \rangle,    
\end{eqnarray}
where $m$ is called as the Trotter number and $\Sigma^{(1)}= \Sigma^{(m+1)} = \Sigma$ which corresponds to periodic boundary condition along the Trotter axis.

\begin{figure}[h]
 \begin{center}
  \includegraphics[scale=0.7]{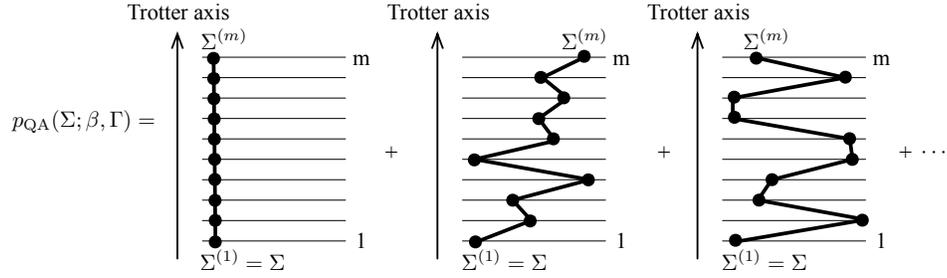}
  \caption{
  Conceptual diagram of path integral.
  In order to calculate the probability of the state $\Sigma$ of quantum system, we add an extra dimension which is called the Trotter axis.
  The probability $p_{\rm QA}(\Sigma;\beta,\Gamma)$ can be calculated by taking sum of configurations of path depicted by the bold line.
  }
  \label{fig:concept_pathintegral}
 \end{center}
\end{figure}

Here we define 
\begin{eqnarray}
 \label{eq:correlation_function}
 &&s(\Sigma^{(j)},\Sigma^{(j+1)}) :=
  \frac{1}{N} \sum_{i=1}^N \delta \left( \sigma_i^{(j)},\sigma_i^{(j+1)}\right),\\
 \label{eq:def_f_beta_gamma}
 &&f(\beta,\Gamma) := N \log \left(1+\frac{Q}{{\rm e}^{\frac{Q\beta\Gamma}{m}}-1} \right),
\end{eqnarray}
where $\sigma_i^{(j)}$ denotes the $i$-th element on the $j$-th Trotter layer.
Then we can obtain
\begin{eqnarray}
 &&\langle \Sigma^{(j)} | {\rm e}^{-\frac{\beta}{m}{\cal H}_{\rm c}} | \Sigma^{(j)'} \rangle
  \propto p_{\rm SA}\left( \Sigma^{(j)};\frac{\beta}{m}\right)
  \delta\left( \Sigma^{(j)},\Sigma^{(j)'}\right),\\
 &&\langle \Sigma^{(j)'} | {\rm e}^{-\frac{\beta}{m}{\cal H}_{\rm q}} | \Sigma^{(j+1)} \rangle
  \propto {\rm e}^{s(\Sigma^{(j)'},\Sigma^{(j+1)}) f(\beta,\Gamma)},
\end{eqnarray}
since ${\cal H}_{\rm c}$ is a diagonal matrix and the equation
\begin{eqnarray}
 \left( \sigma_i^x \right)^l = (\hat{1}_Q - \mathbb{E}_Q)^l = 
  \frac{1}{Q}\left[ (1-Q)^l - 1 \right] \hat{1}_Q + (-1)^l \mathbb{E}_Q
\end{eqnarray}
is satisfied.
Then we obtain 
\begin{eqnarray}
 p_{\rm QA}(\Sigma;\beta,\Gamma) = 
  \frac{1}{Z} \sum_{\Sigma^{(2)}} \cdots \sum_{\Sigma^{(m)}}
  \prod_{j=1}^m p_{\rm SA}(\Sigma^{(j)};\frac{\beta}{m})
  {\rm e}^{s(\Sigma^{(j)},\Sigma^{(j+1)}) f(\beta,\Gamma)}.
\end{eqnarray}
By using path-integral representation, the probability of the state $\Sigma$ of the $d$-dimensional quantum system can be represented by that of $(d+1)$-dimensional classical system approximately (see Fig.\ref{fig:concept_pathintegral}).

\begin{figure}[h]
 \begin{center}
  \includegraphics[scale=0.8]{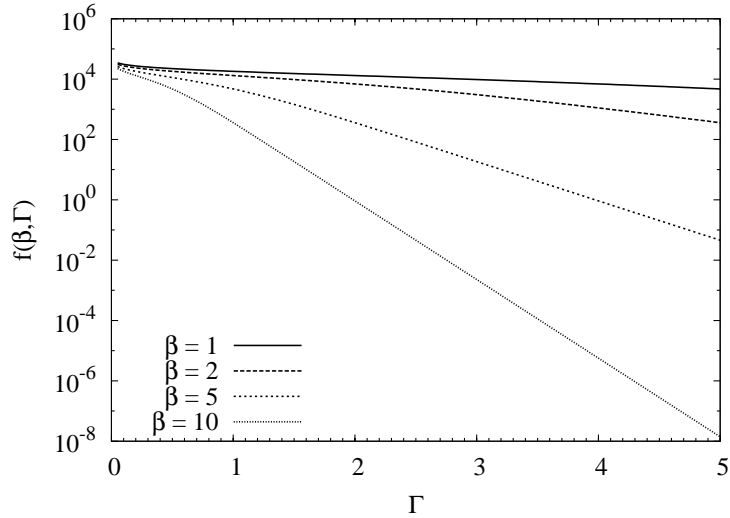}
  \caption{
  $f(\beta,\Gamma)$ as a function of $\Gamma$ for $N=5000$, $Q=30$, and $m=50$.
  }
  \label{fig:graph_f}
 \end{center}
\end{figure}

Here it should be noted that the function $f(\beta,\Gamma)$ is a monotonic decreasing function of inverse temperature $\beta$ and quantum field $\Gamma$ (see Fig.\ref{fig:graph_f}).

In the quantum Monte Carlo simulation by using path-integral representation, we update the state according to the probability as follows:
\begin{eqnarray}
 \label{eq:p_qast_update}
&& p_{\rm QA-ST}^{\rm update}(\sigma_i^{(j)}=s|\Sigma^{(j)} \backslash \sigma_i^{(j)},\Sigma^{(j-1)},\Sigma^{(j+1)};\beta,\Gamma) = \frac{P(s)}{\sum_{s'=1}^Q P(s')},\\
&& P(s) = {\rm e}^{\left\{
 -\frac{\beta}{m}E(\Sigma^{(j)}_s)
 + [ s(\Sigma^{(j-1)},\Sigma^{(j)}_s) + s(\Sigma^{(j)}_s,\Sigma^{(j+1)}) ]
 f(\beta,\Gamma)
\right\}},
\end{eqnarray}
where $\Sigma^{(j)}_s$ represents the state $\sigma_i^{(j)}=s$ given $\Sigma^{(j)}\backslash \sigma_i^{(j)}$.

\section{Some remarks}

In the section 4, we introduced the quantum fluctuation and obtained the probability of the quantum state by the path-integral representation.
Some remarks will be shown in this section.

\subsection{Labels of the clusters}

\begin{figure}[h]
 \begin{center}
  $\begin{array}{ccc}
  \includegraphics[scale=0.8]{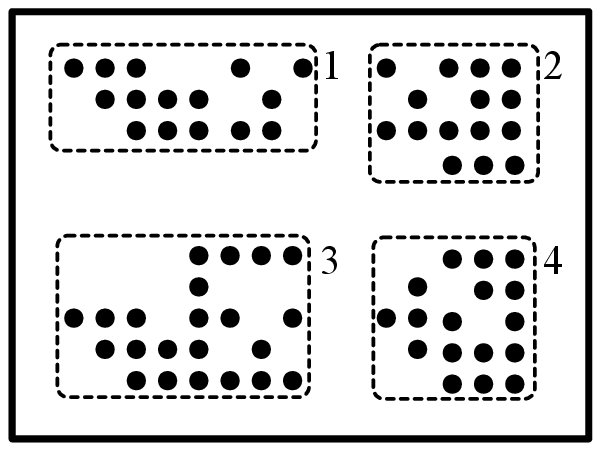}&
   \hspace{5mm}&
  \includegraphics[scale=0.8]{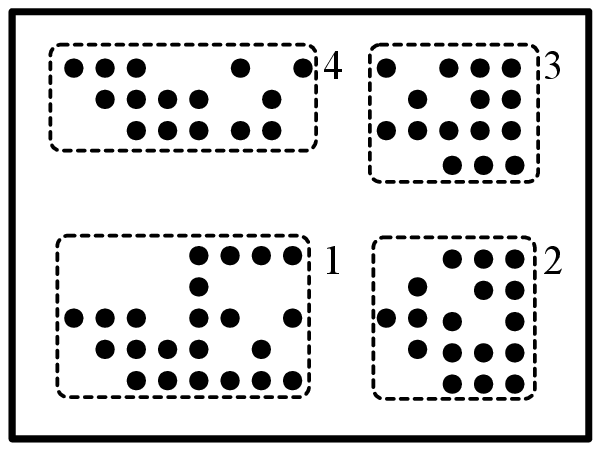} \\
    ({\rm a}) \quad \Sigma^A & \hspace{5mm} & ({\rm b}) \quad \Sigma^B 
  \end{array}$
  \caption{Both (a) and (b) are the same divisions but the name of the clusters are completely different.}
  \label{fig:samedivision}
 \end{center}
\end{figure}

The value $s(\Sigma^{(j)},\Sigma^{(j+1)})$ given by Eq.~(\ref{eq:correlation_function}) expresses the correlation function between the state on the $j$-th Trotter layer and that on the $j+1$-th Trotter layer.
When the divisions are the same but the labels of the clusters are completely different as shown in Fig.~\ref{fig:samedivision}, $s(\Sigma^{(j)},\Sigma^{(j+1)})$ becomes zero.

When we decrease the quantum field $\Gamma$ slow enough, not only division but also the labels of the clusters should become the same.
If the state on the $j$-th layer is $\Sigma^{(j)} = \Sigma_A$ and that on the $j+1$-th layer is $\Sigma^{(j+1)}= \Sigma_B$ such as Fig.~\ref{fig:samedivision} by accident, we cannot only gain a benefit but also are faced with a problem by introducing path-integral representation.
This is similar situation with the domain wall problem in the ferromagnetic Ising model.
In the ferromagnetic Ising model without external magnetic field, the ground state is that all spins are up or down.
Suppose we consider the case that the spins in the left half are up and the spins in the right half are down.
If we execute the standard Monte Carlo simulation, it is difficult to obtain the stable state ({\it i.e.} the ground state) from such an initial state for a short time.
For clustering problems, the domain wall problem gets more seriously comparing with the standard ferromagnetic Ising model since the number of the same divisions is $Q!$.
In order to avoid the domain wall problem, we introduce a new parameter ``modified correlation function'' and approximate the probability distribution.
The definition of the modified correlation function $\tilde{s}(\Sigma^{(j)},\Sigma^{(j+1)})$ is
\begin{eqnarray}
 &&\tilde{s}(\Sigma^{(j)},\Sigma^{(j+1)})
  := \frac{1}{N} \sum_{c=1}^k
  {\rm max}_{c'=1,\cdots,k} [Y(\Sigma^{(j)})Y^T(\Sigma^{(j+1)})]_{c,c'},\\
 &&Y(\Sigma^{(j)}) := (\sigma^{(j)}_1,\sigma^{(j)}_2,\cdots,\sigma^{(j)}_N),
\end{eqnarray}
where $Y(\Sigma^{(j)})$ denotes $N \times Q$ matrix.
For $Q=3$ and $N=2$, $Y(\Sigma_3)$, where $\Sigma_3$ was defined in Table \ref{table:def_sigma_q3n2}, is given as 
\begin{eqnarray}
 Y(\Sigma_3) = 
  \left(
   \begin{array}{cc}
   0 & 1\\
   0 & 0\\
   1 & 0\\
   \end{array}
  \right).
\end{eqnarray}
It should be noted that the correlation function $s(\Sigma^{(j)},\Sigma^{(j+1)})$ can be expressed in a similar way such as
\begin{eqnarray}
 s(\Sigma^{(j)},\Sigma^{(j+1)}) = 
  \frac{1}{N} {\rm Tr} [Y(\Sigma^{(j)})Y^T(\Sigma^{(j+1)})].
\end{eqnarray}

Let us show some properties of the modified correlation function comparing $s(\Sigma^{(j)},\Sigma^{(j+1)})$ in the next section.
Suppose we consider the case for $Q=3$ and $N=7$ in the section 5.1.1 and 5.1.2.

\subsubsection{Example A}

We consider a case that the states on the $j$-th layer and that on the $j+1$-th layer are shown in Table~\ref{table:example1}.

\begin{table}[h]
 \tbl{Example A.}{
  \begin{tabular}{cccccccc}
   \toprule
    & $\sigma_1$ & $\sigma_2$ & $\sigma_3$ & $\sigma_4$ & $\sigma_5$ & $\sigma_6$ & $\sigma_7$ \\
   \hline
   $\Sigma^{(j)}$ & $1$ & $1$ & $1$ & $2$ & $2$ & $3$ & $3$ \\
   $\Sigma^{(j+1)}$ & $3$ & $3$ & $2$ & $3$ & $1$ & $1$ & $2$ \\
   \botrule
  \end{tabular}}
 \label{table:example1}
\end{table}

Here, $Y(\Sigma^{(j)})$ and $Y(\Sigma^{(j+1)})$ are given as
\begin{eqnarray}
 Y(\Sigma^{(j)}) = 
  \left(
   \begin{array}{ccccccc}
    1 & 1 & 1 & 0 & 0 & 0 & 0 \\
    0 & 0 & 0 & 1 & 1 & 0 & 0 \\
    0 & 0 & 0 & 0 & 0 & 1 & 1
   \end{array}
  \right),
  \,\,\,\,\,
  Y(\Sigma^{(j+1)}) = 
  \left(
   \begin{array}{ccccccc}
    0 & 0 & 0 & 0 & 1 & 1 & 0 \\
    0 & 0 & 1 & 0 & 0 & 0 & 1 \\
    1 & 1 & 0 & 1 & 0 & 0 & 0
   \end{array}
  \right).
\end{eqnarray}
Then, the modified correlation function is calculated as
\begin{eqnarray}
 \tilde{s}(\Sigma^{(j)},\Sigma^{(j+1)}) = \frac{1}{7} \sum_{c=1}^3
  {\rm max}_{c'=1,\cdots,3} 
  \left(
   \begin{array}{ccc}
    0 & 1 & 2 \\
    1 & 0 & 1 \\
    1 & 1 & 0
   \end{array}
  \right)_{c,c'} = \frac{4}{7}.
\end{eqnarray}
Note that $s(\Sigma^{(j)},\Sigma^{(j+1)}) = 0$ by the definition.
Next we fix the labels of $\Sigma^{(j)}$ and rename the labels of $\Sigma^{(j+1)}$.
The values of $s(\Sigma^{(j)},{\cal P}_\pi\Sigma^{(j+1)})$ are shown in Table \ref{table:example1_2}, where ${\cal P}_\pi$ ($\pi = 1, \cdots, Q!$) denotes label permutation operator.

\begin{table}[h]
 \tbl{The value of correlation function by applying label permutation operator ${\cal P}_\pi$ for example A.}{
  \begin{tabular}{ccccccccc}
   \toprule
    & $\sigma_1$ & $\sigma_2$ & $\sigma_3$ & $\sigma_4$ & $\sigma_5$ & $\sigma_6$ & $\sigma_7$ & $s(\Sigma^{(j)},{\cal P}_\pi\Sigma^{(j+1)})$ \\
   \hline
   ${\cal P}_1\Sigma^{(j+1)}$ & $3$ & $3$ & $2$ & $3$ & $1$ & $1$ & $2$ & $0$  \\
   ${\cal P}_2\Sigma^{(j+1)}$ & $3$ & $3$ & $1$ & $3$ & $2$ & $2$ & $1$ & $2/7$ \\
   ${\cal P}_3\Sigma^{(j+1)}$ & $1$ & $1$ & $3$ & $1$ & $2$ & $2$ & $3$ & $4/7$ \\
   ${\cal P}_4\Sigma^{(j+1)}$ & $1$ & $1$ & $2$ & $1$ & $3$ & $3$ & $2$ & $3/7$ \\
   ${\cal P}_5\Sigma^{(j+1)}$ & $2$ & $2$ & $3$ & $2$ & $1$ & $1$ & $3$ & $2/7$ \\
   ${\cal P}_6\Sigma^{(j+1)}$ & $2$ & $2$ & $1$ & $2$ & $3$ & $3$ & $1$ & $3/7$ \\
   \botrule
  \end{tabular}}
 \label{table:example1_2}
\end{table}

In this case, the maximum value of $s(\Sigma^{(j)},{\cal P}_\pi\Sigma^{(j+1)})$ is the same as the value of modified correlation function $\tilde{s}(\Sigma^{(j)},\Sigma^{(j+1)})$.

\subsubsection{Example B}

We also consider another case that the states on the $j$-th layer and that on the $j+1$-th layer are shown in Table~\ref{table:example2} as an another example.

\begin{table}[h]
 \tbl{Example B.}{
  \begin{tabular}{cccccccc}
   \toprule
    & $\sigma_1$ & $\sigma_2$ & $\sigma_3$ & $\sigma_4$ & $\sigma_5$ & $\sigma_6$ & $\sigma_7$ \\
   \hline
   $\Sigma^{(j)}$ & $1$ & $2$ & $2$ & $2$ & $2$ & $3$ & $3$ \\
   $\Sigma^{(j+1)}$ & $2$ & $1$ & $1$ & $1$ & $3$ & $1$ & $1$ \\
   \botrule
  \end{tabular}}
 \label{table:example2}
\end{table}

Here, $Y(\Sigma^{(j)})$ and $Y(\Sigma^{(j+1)})$ are given as
\begin{eqnarray}
 Y(\Sigma^{(j)}) = 
  \left(
   \begin{array}{ccccccc}
    1 & 0 & 0 & 0 & 0 & 0 & 0 \\
    0 & 1 & 1 & 1 & 1 & 0 & 0 \\
    0 & 0 & 0 & 0 & 0 & 1 & 1
   \end{array}
  \right),
  \,\,\,\,\,
  Y(\Sigma^{(j+1)}) = 
  \left(
   \begin{array}{ccccccc}
    0 & 1 & 1 & 1 & 0 & 1 & 1 \\
    1 & 0 & 0 & 0 & 0 & 0 & 0 \\
    0 & 0 & 0 & 0 & 1 & 0 & 0
   \end{array}
  \right).
\end{eqnarray}
Then, the modified correlation function is calculated as
\begin{eqnarray}
 \tilde{s}(\Sigma^{(j)},\Sigma^{(j+1)}) = \frac{1}{7} \sum_{c=1}^3
  {\rm max}_{c'=1,\cdots,3} 
  \left(
   \begin{array}{ccc}
    0 & 1 & 0 \\
    3 & 0 & 1 \\
    2 & 1 & 1
   \end{array}
  \right)_{c,c'} = \frac{6}{7}.
\end{eqnarray}
Note that $s(\Sigma^{(j)},\Sigma^{(j+1)}) = 0$ by the definition.
As in the section 5.1.1, when we fix the labels of $\Sigma^{(j)}$ and rename the labels of $\Sigma^{(j+1)}$, the values of $s(\Sigma^{(j)},{\cal P}_\pi\Sigma^{(j+1)})$ are shown in Table \ref{table:example2_2}.

\begin{table}[h]
 \tbl{The value of correlation function by applying label permutation operator ${\cal P}_\pi$ for example B.}{
  \begin{tabular}{ccccccccc}
   \toprule
   & $\sigma_1$ & $\sigma_2$ & $\sigma_3$ & $\sigma_4$ & $\sigma_5$ & $\sigma_6$ & $\sigma_7$ & $s(\Sigma^{(j)},{\cal P}_\pi\Sigma^{(j+1)})$ \\
   \hline
   ${\cal P}_1\Sigma^{(j+1)}$ & $2$ & $1$ & $1$ & $1$ & $3$ & $1$ & $1$ & $0$ \\
   ${\cal P}_2\Sigma^{(j+1)}$ & $3$ & $1$ & $1$ & $1$ & $2$ & $1$ & $1$ & $1/7$ \\
   ${\cal P}_3\Sigma^{(j+1)}$ & $1$ & $2$ & $2$ & $2$ & $3$ & $2$ & $2$ & $4/7$ \\
   ${\cal P}_4\Sigma^{(j+1)}$ & $3$ & $2$ & $2$ & $2$ & $1$ & $2$ & $2$ & $3/7$ \\
   ${\cal P}_5\Sigma^{(j+1)}$ & $1$ & $3$ & $3$ & $3$ & $2$ & $3$ & $3$ & $4/7$ \\
   ${\cal P}_6\Sigma^{(j+1)}$ & $2$ & $3$ & $3$ & $3$ & $1$ & $3$ & $3$ & $2/7$ \\
   \botrule
  \end{tabular}}
  \label{table:example2_2}
\end{table}

In this case, the maximum value of $s(\Sigma^{(j)},{\cal P}_\pi\Sigma^{(j+1)})$ is not the same as the modified correlation function $\tilde{s}(\Sigma^{(j)},\Sigma^{(j+1)})$.
Although the modified correlation function expresses some kinds of similarity between the states $\Sigma^{(j)}$ and $\Sigma^{(j+1)}$, it is not necessary commutative {\it i.e.} $\tilde{s}(\Sigma^{(j)},\Sigma^{(j+1)}) \neq \tilde{s}(\Sigma^{(j+1)},\Sigma^{(j)})$.
In the case of example A,
\begin{eqnarray}
 \tilde{s}(\Sigma^{(j)},\Sigma^{(j+1)}) = \tilde{s}(\Sigma^{(j+1)},\Sigma^{(j)}) = \frac{4}{7}.
\end{eqnarray}
On the other hand, in the case of example B, 
\begin{eqnarray}
 \tilde{s}(\Sigma^{(j)},\Sigma^{(j+1)}) = \frac{6}{7},
  \,\,\,\,\,
  \tilde{s}(\Sigma^{(j+1)},\Sigma^{(j)}) = \frac{5}{7}.
\end{eqnarray}
Here we summarize the properties of the modified correlation function:
\begin{itemize}
 \item The modified correlation function does not necessary commute.
       \begin{eqnarray}
	\tilde{s}(\Sigma^{(j)},\Sigma^{(j+1)}) \neq \tilde{s}(\Sigma^{(j+1)},\Sigma^{(j)}).
       \end{eqnarray}
 \item For any ${\cal P}_\pi$ ($\pi=1,\cdots,Q!$), the inequality
       \begin{eqnarray}
	\nonumber
	0 \le s(\Sigma^{(j)},{\cal P}_\pi\Sigma^{(j+1)}) 
	 = {\rm Tr} \left[ Y(\Sigma^{(j)})Y({\cal P}_\pi\Sigma^{(j+1)})^T \right]
	 \le \tilde{s}(\Sigma^{(j)},\Sigma^{(j+1)}) \le 1
       \end{eqnarray}
       is satisfied.
\end{itemize}
Although the modified correlation function is not necessary the same as the maximum value of $s(\Sigma^{(j)},{\cal P}_\pi\Sigma^{(j+1)})$, we use this value in our proposed algorithm.
This is because our aim is to find not so bad solution of given problem as mentioned before.

\subsection{Speed-up by Using Modified Correlation Function}

When we adopt the quantum Monte Carlo simulation by the path-integral representation, it is enough to update the state according to Eq.~(\ref{eq:p_qast_update}) in principle.
In the previous section, we introduced the new parameter ``modified correlation function'' to avoid the so-called domain wall problem.
Instead of Eq.~(\ref{eq:p_qast_update}), we adopt the following the transition probability based on modified correlation function:
\begin{eqnarray}
\nonumber 
&&p_{\rm QA-ST+modify}^{\rm update}(\sigma_i^{(j)}=s|\Sigma^{(j)}\backslash \sigma_i^{(j)},\Sigma^{(j-1)},\Sigma^{(j+1)};\beta,\Gamma)\\
\label{eq:prob_qastpurity} 
&&=\frac
  {
  \exp\left[
       -\frac{\beta}{m}E(\Sigma^{(j)}_s)+\tilde{S}(\Sigma^{(j-1)},\Sigma^{(j)}_s,\Sigma^{(j+1)})f(\beta,\Gamma)
      \right]
}{
\sum_{t=1}^Q \exp\left[
       -\frac{\beta}{m}E(\Sigma^{(j)}_t)+\tilde{S}(\Sigma^{(j-1)},\Sigma^{(j)}_t,\Sigma^{(j+1)})f(\beta,\Gamma)
      \right]
},\\
 &&\tilde{S}(\Sigma^{(j-1)},\Sigma^{(j)}_s,\Sigma^{(j+1)})
  := \tilde{s}(\Sigma^{(j-1)},\Sigma^{(j)}_s) + \tilde{s}(\Sigma^{(j)}_s,\Sigma^{(j+1)}).
\end{eqnarray}
Since $\tilde{s}(\Sigma^{(j-1)},\Sigma^{(j)})$ is not necessary commutative, there are four possibilities of the definition of $\tilde{S}(\Sigma^{(j-1)},\Sigma^{(j)},\Sigma^{(j+1)})$ as follows:
\begin{eqnarray}
 \label{eq:def_purity_oneoffour}
 &&\tilde{S}(\Sigma^{(j-1)},\Sigma^{(j)}_s,\Sigma^{(j+1)})
  = \tilde{s}(\Sigma^{(j-1)},\Sigma^{(j)}_s) + \tilde{s}(\Sigma^{(j)}_s,\Sigma^{(j+1)}),\\
 &&\tilde{S}(\Sigma^{(j-1)},\Sigma^{(j)}_s,\Sigma^{(j+1)})
  = \tilde{s}(\Sigma^{(j-1)},\Sigma^{(j)}_s) + \tilde{s}(\Sigma^{(j+1)},\Sigma^{(j)}_s),\\
 &&\tilde{S}(\Sigma^{(j-1)},\Sigma^{(j)}_s,\Sigma^{(j+1)})
  = \tilde{s}(\Sigma^{(j)}_s,\Sigma^{(j-1)}) + \tilde{s}(\Sigma^{(j)}_s,\Sigma^{(j+1)}),\\
 &&\tilde{S}(\Sigma^{(j-1)},\Sigma^{(j)}_s,\Sigma^{(j+1)})
  = \tilde{s}(\Sigma^{(j)}_s,\Sigma^{(j-1)}) + \tilde{s}(\Sigma^{(j+1)},\Sigma^{(j)}_s).
\end{eqnarray}
For small systems we confirm that when we adopt Eq.~(\ref{eq:def_purity_oneoffour}) as the definition of $\tilde{S}(\Sigma^{(j-1)},\Sigma^{(j)}_s,\Sigma^{(j+1)})$, we can obtain better solution than the other choices.
Then, we adopt the relation given by Eq.~(\ref{eq:def_purity_oneoffour}) as the definition of $\tilde{S}(\Sigma^{(j-1)},\Sigma^{(j)}_s,\Sigma^{(j+1)})$.

\subsection{Simultaneous Control of Thermal Fluctuation and Quantum Fluctuation}

In our algorithm given by Eq.~(\ref{eq:prob_qastpurity}), thermal fluctuation $\beta/m$ and quantum fluctuation $f(\beta,\Gamma)$ coexist.
Then, it is expected that we can obtain the better solution than simulated annealing by controlling the thermal fluctuation and quantum fluctuation simultaneously with ingenuity.

\begin{figure}[h]
 \begin{center}
  \includegraphics[scale=0.5]{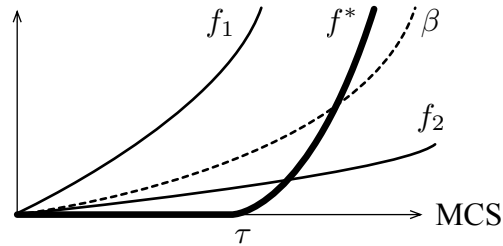}
  \caption{
  The dotted curve denotes the schedule of inverse temperature.
  $f_1$ is too rapid for decreasing the quantum field.
  On the other hand, $f_2$ is too slow for decreasing the quantum field.
  $f^*$ denotes the best schedule.
  }
  \label{fig:proposed_algorithm}
 \end{center}
\end{figure}

In order to investigate how to control both of the thermal fluctuation and quantum fluctuation, we first consider limiting cases.
When the temperature is very low than the quantum field, {\it i.e.} $\beta/m \gg f(\beta,\Gamma)$, the probability distributions in each layer obey Eq.~(\ref{eq:p_sa}).
On the other hand, if the quantum field is very weak, {\it i.e.} $\beta/m \ll f(\beta,\Gamma)$, the state in each layers becomes the same state such as $\Sigma = \Sigma^{(j)}$ for all $j$.
By calculating small systems, we find how to control the thermal fluctuation and quantum fluctuation simultaneously.
The best condition is as follows: 
(i) At the first step, to find the metastable state on each layers, we set the temperature $\beta/m \gg f(\beta,\Gamma)$. 
(ii)After that, $f(\beta,\Gamma)$ overstrides thermal fluctuation.
The curve $f^*$ in Fig.~\ref{fig:proposed_algorithm} shows a conceptual diagram for our proposed schedule.

The dotted curve in Fig.~\ref{fig:proposed_algorithm} indicates the schedule of cooling temperature which corresponds to the simulated annealing.
The curve depicted $f^*$ in Fig.~\ref{fig:proposed_algorithm} is the best schedule for obtaining better solution than the simulated annealing.
In the schedule depicted $f_1$ in Fig.~\ref{fig:proposed_algorithm}, quantum mixing effect does not make sense.
In the schedule depicted $f_2$ in Fig.~\ref{fig:proposed_algorithm}, on the other hand, the states on each layers behave independently.
The quantum fluctuation effect is strong than the thermal fluctuation effect in the schedule depicted $f_2$.
It is essentially the same as the simulated annealing.

Here we assume the scheduling functions of temperature and quantum field as follows:
\begin{eqnarray}
 &&\beta(t) = \beta_0 r_\beta^t,\\
 &&\Gamma(t) = \infty,  \,\,\,\,\, (t < \tau),  \,\,\,\,\,\,\,\,
  \Gamma(t) = \Gamma_0 \exp(-r_\Gamma^{t-\tau}) \,\,\,\,\, (t \ge \tau),
\end{eqnarray}
where $\tau$ corresponds to the time $\beta(\tau) = m$.
When $Q\beta\Gamma/m \ll 1$, the interaction along the Trotter axis $f(\beta,\Gamma)$ given by Eq.~(\ref{eq:def_f_beta_gamma}) is approximately given as
\begin{eqnarray}
 f(\beta,\Gamma) \sim -N \log(\frac{\beta\Gamma}{m}) 
  = N r_\Gamma^t - N \log (\frac{\beta\Gamma_0}{m}).
\end{eqnarray}
From this equation, it is enough to set large enough $\Gamma_0$ and $r_\beta < r_\Gamma$ in order to prepare the schedule $f^*$.
In other words, independent simulated annealing is performed until $\tau$ and after that we decrease quantum field to obtain the better solution than the conventional simulated annealing.

\section{Results}\label{sec:results}

We perform numerical experiment for the following three problems by proposed quantum annealing and simulated annealing for comparison.
We prepare the number of replicas $m = 50$ in the whole experiments.
The initial inverse temperature and the initial quantum field are set to be $\beta_0 = 0.2m$ and $\Gamma_0 = {\rm e}^{1/2}$, respectively. 
Here the ratio of changing temperature is set to be $r_\beta = 1.05$.
In the simulated annealing, we prepare independent $55$ samples and use the same initial temperature and the ratio of changing temperature as the case of quantum annealing.
We study three problems as follows:
\begin{description}
 \item[(a)] Evaluation of mixture of Gaussian by using MNIST data\cite{MNIST} \\
	    ($Q=30$,$N=5000$)
 \item[(b)] Evaluation of latent Dirichlet allocation\cite{Blei-2003} by using Reuters data\cite{REUTERS} \\
	    ($Q=20$,$N=2000$)
 \item[(c)] Evaluation of latent Dirichlet allocation by using NIPS corpus\cite{NIPS} \\
	    ($Q=20$,$N=1000$)
\end{description}
In all of the experiments, we fix the ratio of changing the inverse temperature $r_\beta = 1.05$.

\begin{figure}[h]
 \begin{center}
  $\begin{array}{ccccc}
   \includegraphics[scale=0.4]{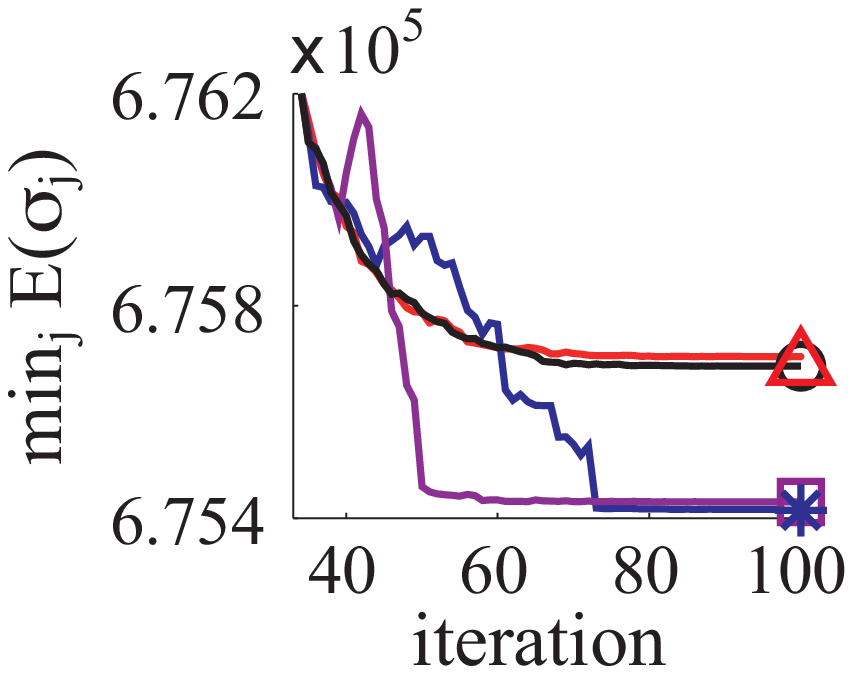}&
    \hspace{3mm}&
   \includegraphics[scale=0.4]{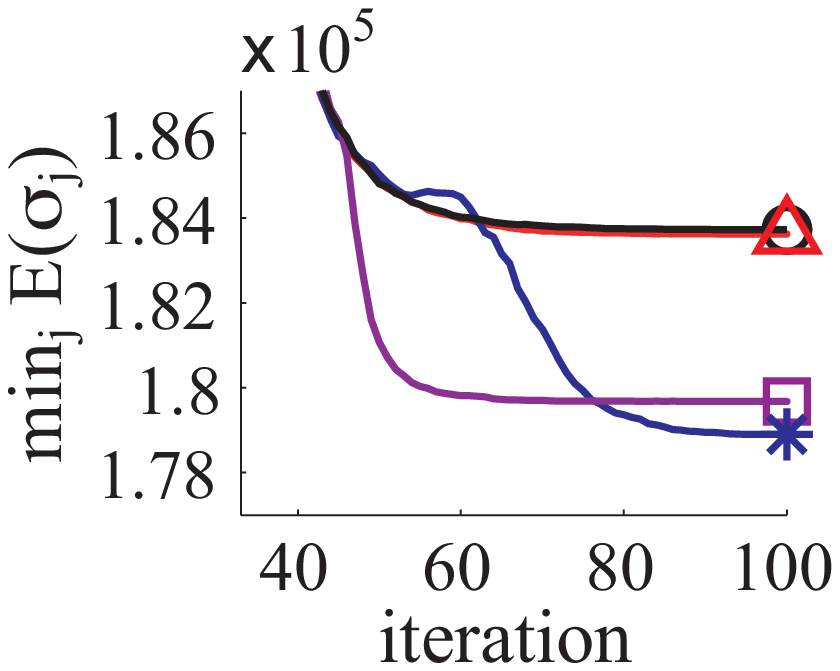}&
   \hspace{3mm}&
   \includegraphics[scale=0.4]{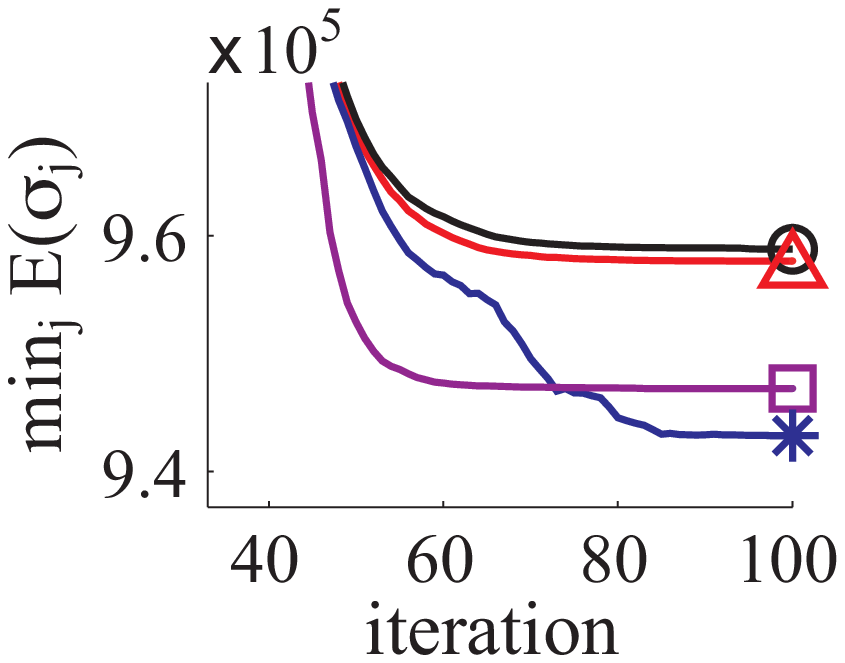}\\
    ({\rm a}) & \hspace{3mm} & ({\rm b}) & \hspace{3mm} & ({\rm c})
   \end{array}$
  \caption{
  Time development of the minimum energy in all of the layers. 
  The circles represent results obtained by simulated annealing for comparison.
  The triangles, asterisks, and squares denote obtained results by proposed quantum annealing for $r_\Gamma = 1.02$, $r_\Gamma = 1.10$, and $r_\Gamma = 1.20$, respectively.
  (a) Evaluation of mixture of Gaussian by using MNIST data ($Q=30$,$N=5000$).
  (b) Evaluation of latent Dirichlet allocation by using Reuters data ($Q=20$,$N=2000$).
  (c) Evaluation of latent Dirichlet allocation by using NIPS corpus ($Q=20$,$N=1000$).
  }
  \label{fig:results}
 \end{center}
\end{figure}

Figure \ref{fig:results} denotes the time development of the minimum energy in all of the layers for each problems.
The circles in Fig.~\ref{fig:results} represent results obtained by simulated annealing for comparison.
The triangles, asterisks, and squares denote obtained results by proposed quantum annealing for $r_\Gamma = 1.02$, $r_\Gamma = 1.10$, and $r_\Gamma = 1.20$, respectively.
The schedule for $r_\Gamma = 1.02$ corresponds to schedule depicted $f_2$ in Fig.~\ref{fig:proposed_algorithm} whereas the schedules for $r_\Gamma = 1.10$ and $r_\Gamma = 1.20$ correspond to schedule depicted $f^*$ in Fig.~\ref{fig:proposed_algorithm}.
From Fig.~\ref{fig:results} it is clear that we can obtain the better solution than simulated annealing if we adopt the schedule $f^*$.

\section{Conclusion}\label{sec:conclusion}

In this paper, we developed hybrid type of the quantum annealing for clustering problems.
When we apply the standard quantum annealing for these problems, we face on the difficulty of the domain wall problem.
Since the number of the same division is $Q!$, the domain wall problem gets more serious for large $Q$.
To avoid such a problem, we introduced new parameter ``modified correlation function'' instead of the standard correlation function.
We investigated the best schedule of changing the thermal fluctuation and the quantum fluctuation simultaneously.
We first apply strong quantum field and then we obtain the metastable states.
At the second step, we can obtain the better solution than the simulated annealing by decreasing the quantum field.
Actually, we succeeded to obtain the better solution for clustering problems than the simulated annealing.
We expect that the proposed schedule of changing the thermal fluctuation and the quantum fluctuation is generally efficient for other type of problems.
However, it is an open problem when to use the quantum annealing.
To solve this problem, we should study the efficiency of quantum annealing for the problems where the {\it difficulty} of the problem can be controlled.

The authors are grateful to Naoki Kawashima, Seiji Miyashita, Hiroshi Nakagawa, Daiji Suzuki, and member of T-PRIMAL for their valuable comments.
S.T. is partly supported by Grant-in-Aid for Young Scientists Start-up (21840021) from the JSPS, MEXT Grant-in-Aid for Scientific Research (B) (22340111), and the ``Open Research Center'' Project for Private Universities: matching fund subsidy from MEXT.
R.T. is partly supported by Global COE Program ``the Physical Sciences Frontier'', MEXT, Japan. 
The computation in the present work was performed on computers at the Supercomputer Center, 
Institute for Solid State Physics and University of Tokyo and at Taisuke Sato's group, Tokyo Institute of Technology.

\appendix{Implementation Method of New Kind of Fluctuation}

In this paper, we studied a quantum effect on clustering problems which are expressed by the Potts model.
Relationship between a phase transition and a performance of quantum annealing is very important, although we did not mention it in this paper.
Many researchers have studied the order of the phase transition of a given problem and concluded whether the quantum annealing is efficient or not for the problems\cite{Charkrabarti-book1,Charkrabarti-book2,Caneva-2007,Suzuki-2009,Young-2010}.
It is one of main topics on quantum annealing.

Suppose there is a transition point at halfway of a control parameter such as magnetic field.
If a first-order phase transition occurs, the quantum adiabatic computation, in principle, does not obtain a good solution because of a level-crossing problem.
There is a trivial example of level-crossing problem: the ferromagnetic Ising model with longitudinal magnetic field $h^z$.
In this model, level crossing occurs at $h^z=0$.
The initial state is set to be a ground state of negative $h^z$.
When we sweep $h^z$ from negative to positive, the state cannot follow the adiabatic limit of the state at all.
If a second-order phase transition occurs, on the other hand, the growth of correlation length does not follow for finite speed of sweeping of control parameter.
The equilibrium value of correlation length diverges at the second-order phase transition point.
In practice, the quantum annealing does not succeed to obtain the best solution for systems in which the second-order phase transition occurs.
If there is no phase transition, the quantum annealing is expected to find the best solution.

Our purposes are to control the order of phase transition and, what is more, to erase the phase transition by adding some kind of fluctuation.
Recently, we constructed a model in which the order of the phase transition can be changed by controlling a fluctuation\cite{Tamura-2010,Tanaka-2010,Tanaka-2011}.
This model is called the Potts model with invisible states.
We found that a first-order phase transition is driven by the effect of invisible states (invisible fluctuation) in the ferromagnetic Potts model with invisible states.
Although the invisible fluctuation itself seems to be getting worse for quantum annealing from the above discussion, the invisible fluctuation changes the order of the phase transition without changing an essence of problems.
We expect that there are fluctuations which wipe a phase transition.
Thus, to introduce new kinds of fluctuation is important for optimization problems.
In this section, we introduce the concept of invisible fluctuation.

We first consider the standard Potts model.
The Hamiltonian of this model is given by
\begin{eqnarray}
 \label{eq:standard_potts}
 {\cal H}_{\rm standard} = 
  \sum_{\langle i,j \rangle \in E(G)} J_{ij} \delta_{\sigma_i,\sigma_j},
  \qquad
  \sigma_i = 1, \cdots, Q,
\end{eqnarray}
where $E(G)$ denotes the set of edges of given graph $G$.
Eq.~(\ref{eq:standard_potts}) is called the $Q$-state Potts model.
Here we assume $Q$ is a natural number.
Suppose we consider the ferromagnetic case {\it i.e.} $J_{ij} = -J$ for all $\langle i,j \rangle \in E(G)$.
A second-order phase transition occurs when $Q \le 4$ whereas a first-order phase transition occurs when $Q >4$ on two-dimensional lattice.
It is interesting that the order of the phase transition of the standard ferromagnetic Potts model can be changed by the number of states $Q$.
The ground state of this model is that all of the spins have the same value.
The number of ground states is $Q$.
Then, the phase transition accompanies spontaneous $Q$-fold symmetry breaking.
The standard Potts model has been regarded as the standard model not only in statistical physics but also in wide area of science.

We consider the Hamiltonian of the Potts model with invisible states as follows:
\begin{eqnarray}
 \label{eq:invpotts}
 {\cal H}_{\rm inv} = \sum_{\langle i,j \rangle \in E(G)} J_{ij} \delta_{\sigma_i,\sigma_j}
  \sum_{\alpha=1}^{Q} \delta_{\sigma_i,\alpha},
  \qquad
 \sigma_i = 1, \cdots, Q+R. 
\end{eqnarray}
This model is called the ($Q$,$R$)-state Potts model\cite{Tamura-2010,Tanaka-2010,Tanaka-2011}.
Suppose we consider the case for $J_{ij}=-J$ for all $\langle i,j \rangle \in E(G)$ for simplicity as the previous example.
If and only if $1 \le \sigma_i = \sigma_j \le Q$, interaction $-J$ works.
Thus, the number of ground states is $Q$.
Note that if $R=0$, this model is equivalent to the standard ferromagnetic Potts model.
Hereafter we call the states where $1 \le \sigma_i \le Q$ ``colored states'' whereas the states where $Q+1 \le \sigma_i \le Q+R$ ``invisible states''.

Here we consider two spin system.
The number of excited states of the standard ferromagnetic Potts model given by Eq.~(\ref{eq:standard_potts}) is $Q^2-Q$.
On the other hand, the number of excited states of the ($Q$,$R$)-state Potts model given by Eq.~(\ref{eq:invpotts}) is $Q^2-Q+2QR+R^2$.
The ($Q$,$R$)-state Potts model does not change the number of degeneracy of the ground states.
However the number of excited states are different.
As the number of sites $N$ increases, since density of state changes, it is expected that nature of phase transition changes. 

The order parameter of the ($Q$,$R$)-state Potts model is defined as
\begin{eqnarray}
 {\bf m} = \frac{1}{N} \sum_{i=1}^N {\bf e}^{\sigma_i},
\end{eqnarray}
where ${\bf e}^\alpha$ ($\alpha = 1,\cdots,Q$) represents $Q$ unit vectors pointing in the $Q$ symmetric direction of a hypertetrahedron in $Q-1$ dimensions.
It should be noted that the Kronecker's delta can be represented by using ${\bf e}^\alpha$ as follows:
\begin{eqnarray}
 \delta_{\alpha,\beta} = \frac{1+(Q-1){\bf e}^\alpha \cdot {\bf e}^\beta}{Q}.
\end{eqnarray}
The definition of order parameter is the same as that of the standard ferromagnetic $Q$-state Potts model.
The phase transition accompanies $Q$-fold symmetry if a phase transition takes place in this model.
We investigated this model by mean-field analysis and Monte Carlo simulation\cite{Tamura-2010,Tanaka-2010,Tanaka-2011}.
In these papers, we concluded that the invisible states drive the first-order phase transition and a phase transition with $Q$-fold symmetry breaking occurs at finite temperature.

Before concluding this section, we discuss why a first-order phase transition is driven by the invisible fluctuation.
The Hamiltonian given in Eq.~(\ref{eq:invpotts}) can be transformed exactly by comparing the partition function as follows:
\begin{eqnarray}
 \label{eq:invpotts_eff}
 &&{\cal H}_{\rm inv}^{\rm eff} = 
  \sum_{\langle i,j \rangle \in E(G)} J_{ij} \delta_{\tau_i,\tau_j}
  \sum_{\alpha=1}^Q \delta_{\tau_i,\alpha}
  - T \log r \sum_{i=1}^N \delta_{\tau_i,0},\\
  &&\tau_i = 0,1,\cdots,Q,
\end{eqnarray}
where $T$ represents a temperature.
Here we rename the label of the invisible states from $Q+1 \le \sigma_i \le Q+R$ to $\tau_i = 0$.
The second term means chemical potential of the invisible states.
The ($Q$,$R$)-state Potts model can be mapped onto the annealed diluted Potts model whose chemical potential depends on temperature linearly.
As we change temperature, the chemical potential is varied.
This concept is similar with our hybrid quantum annealing method.
It should be noted that temperature-dependency of the chemical potential comes from the number of invisible states, in other words, entropy of the invisible states.

As mentioned above, the invisible fluctuation itself is inefficient for optimization problem.
However the quantum annealing expects to be a powerful method by adding a new fluctuation -- quantum fluctuation.
In a similar way, it is possible that there is a ``good'' fluctuation for optimization problems.
In this section, we have considered the effect of invisible fluctuation.
The invisible fluctuation is one of ``entropic fluctuation''.
The order of phase transition is decided by the density of states.
Then, such a entropic fluctuation is expected to wipe a phase transition.
We believe that an entropic fluctuation which is constructed as the invisible fluctuation makes some advantages for optimization problems.

\bibliographystyle{ws-procs9x6}
\bibliography{ws-pro-sample}

\end{document}